\documentclass[fleqn,usenatbib]{mnras}
\pdfoutput=1

\usepackage{longtable,lscape}
\usepackage{newtxtext,newtxmath}
\usepackage[T1]{fontenc}
\usepackage{ae,aecompl}
\usepackage{graphicx}	
\usepackage{amsmath}	
\usepackage{amssymb}	

\title[Near-IR spectra of MC candidate high-mass YSOs]{Unveiling the nature of candidate high-mass young stellar objects in the Magellanic Clouds with near-IR spectroscopy} 

\author[M.\ Reiter et al.]{
Megan Reiter,$^{1,2}$\thanks{e-mail: megan.reiter@stfc.ac.uk (MR)}
Omnarayani Nayak,$^{3}$
Margaret Meixner,$^{3,4}$
Olivia Jones$^{2}$
\\
$^{1}$ Department of Astronomy, University of Michigan, 311 West Hall, 1085 South University Avenue, Ann Arbor, MI 48109, USA \\
$^{2}$ UK Astronomy Technology Centre, Royal Observatory, Blackford Hill, Edinburgh, EH9 3HJ, UK \\
$^{3}$ Department of Physics \& Astronomy, Johns Hopkins University, 3400 N. Charles St., Baltimore, MD 21218, USA \\
$^{4}$ Space Telescope Science Institute, 3700 San Martin Drive, Baltimore, MD 21218, USA
}

\date{Accepted XXX. Received YYY; in original form ZZZ}

\pubyear{2018}

\begin{document}

\label{firstpage}
\pagerange{\pageref{firstpage}--\pageref{lastpage}}
\maketitle


\begin{abstract}

As nearby neighbors to the Milky Way, the Large and Small Magellanic Clouds (LMC and SMC) provide a unique opportunity to study star formation in the context of their galactic ecosystems. 
Thousands of young stellar objects (YSOs) have been characterized with large-scale \emph{Spitzer} and \emph{Herschel} surveys. 
In this paper, we present new near-IR spectroscopy of five high-mass YSOs in the LMC and one in the SMC.
We detect multiple hydrogen recombination lines, as well as He~{\sc i}~2.058~$\mu$m, H$_2$, [Fe~{\sc ii}], and [S~{\sc iii}] in these highly excited sources. 
We estimate the internal extinction of each source and find that it is highest for sources with the youngest evolutionary classifications.
Using line ratios, we assess the dominant excitation mechanism in the three sources where we detect both H$_2$~2.12~$\mu$m and [Fe~{\sc ii}]~1.64~$\mu$m.
In each case, photoexcitation dominates over shock excitation.
Finally, we detect CO bandhead \emph{absorption} in one of our LMC sources.
While this feature is often associated with evolved stars, this object is likely young with strong PAH and fine-structure emission lines tracing an H~{\sc ii} region detected at longer wavelengths. 
Compared to high-mass YSOs in the Galaxy, our sources have higher bolometric and line luminosities, consistent with their selection as some of the brightest sources in the LMC and SMC.
\end{abstract}

\begin{keywords}
stars: formation – stars: massive – stars: pre-main-sequence – stars: protostars
\end{keywords}


\section{Introduction}\label{s:intro}

The formation and evolution of high-mass stars remains an open question in astrophysics. 
The recent explosion of Galactic plane surveys has provided an unprecedented view of thousands of actively star-forming clouds throughout the plane of the Milky Way \citep[e.g.,][]{hoare2005,urquhart2008,schuller2009,molinari2010,aguirre2011}. 
Unlike previous, targeted observations, these surveys did not select for a particular star-formation indicator, instead providing an unbiased sample. 
This large population of clumps has enabled construction of an evolutionary sequence and mass function for high-mass stars \citep[e.g,][]{svoboda2016}.

Galaxy-wide surveys, particularly those that target galaxies external to the Milky Way, allow detailed studies of the association between star formation and environment. 
Two of the closest galaxies to the Milky Way, the Large and Small Magellanic Clouds (LMC and SMC), are particularly prime for observational scrutiny  \citep[e.g.,][]{whitney2008,dobbs2010,louie2013,sewilo2013,choi2015,lewis2015,ochsendorf2016}. 
The LMC and SMC provide many sources at a uniform distance ($\sim 50$~kpc, \citealt{pietrzynski2013} and $\sim 60$~kpc, \citealt{scowcroft2016}, respectively) in irregular galaxies with favorable orientations. 
Unlike other extragalactic laboratories, 
the LMC and SMC are close enough to resolve individual high-mass star-forming regions. 
Multi-wavelength surveys particularly with \emph{Spitzer} and \emph{Herschel} have uncovered a wealth of candidate young stellar objects (YSOs) in both galaxies \citep[e.g.,][]{meixner2006,whitney2008,meixner2013,sewilo2013,seale2014}.
Hundred of candidate YSOs have been targeted for follow-up observations with the \emph{Spitzer InfraRed Spectrograph} \citep[\emph{Spitzer IRS};][]{oliveira2009,seale2009,oliveira2011,woods2011,oliveira2013,ruffle2015,jones2017}.  
Spectroscopic data reveal 
absorption due to molecular ices and silicate dust in the youngest protostars 
while more evolved sources show fine-structure emission lines and UV-pumped polycyclic aromatic hydrocarbons (PAHs) that are typically excited in H~{\sc ii} regions.

New observations with the Atacama Large Millimeter/sub-millimeter Array (ALMA) provide maps of the continuum and molecular emission with comparable sensitivity and resolution to observations of Galactic clouds.
For the first time, it is possible to resolve individual sites of high-mass star formation within the structured natal gas that characterizes the earliest stages. 
Recent ALMA observations of multiple high-mass star-forming regions in the LMC show protostars forming in dense, filamentary molecular gas with complex dynamics \citep[e.g.,][]{indebetouw2013,fukui2015,nayak2016,saigo2017,wong2017,nayak2018,naslim2018}.

The most complete view of high-mass star formation in the LMC and SMC requires multi-wavelength observations to construct an evolutionary picture. 
In this paper, we present ground-based near-IR spectroscopy of five high-mass star-forming regions in the LMC and one in the SMC. 
We sample high-mass, high-luminosity sources, including 
the super star cluster (SSC) candidate H72.97-69.39, which was recently identified as a younger analog of 30~Doradus \citep[30~Dor, see][Nayak et al.\ submitted]{och17}.
High-sensitivity near-IR spectroscopy with NIRSpec on the \emph{James Webb Space Telescope (JWST)} will enable a more complete spectroscopic census of a broader range of sources. 
When combined with on-going surveys with other observatories like SOFIA and ALMA, this provides a powerful opportunity to constrain high-mass star-formation at low metallicities.


\section{Target Selection}\label{s:targets}

\begin{figure*}
    \includegraphics[angle=0,scale=0.375]{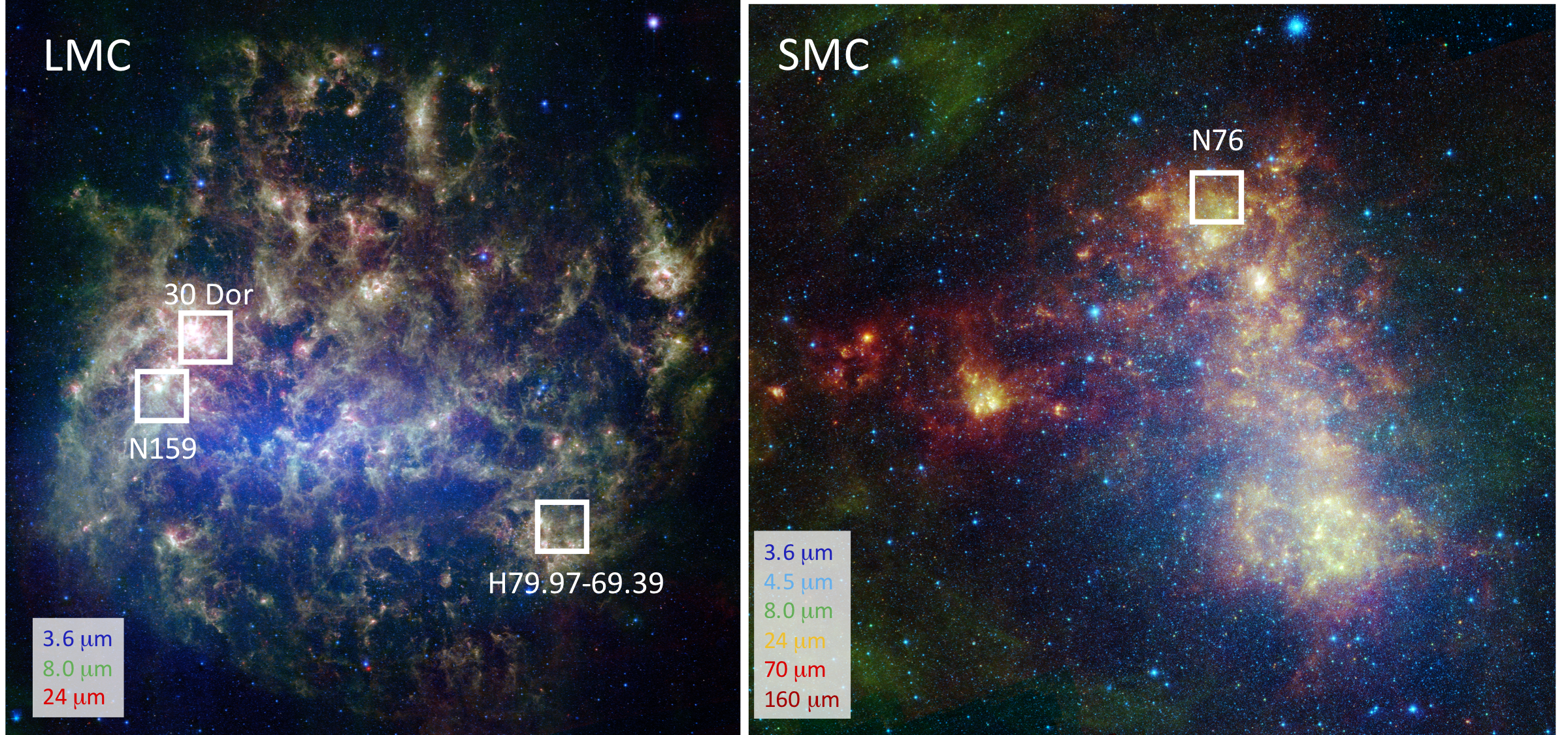} 
    \caption{Approximate location of the YSOs targeted in this program shown on three-color \emph{Spitzer} images of the Large \citep[\textit{left};][]{meixner2006} and Small \citep[\textit{right};][]{gordon2011} Magellanic Clouds. 
      White boxes and labels indicate the regions targeted for near-IR spectroscopy. 
      Most regions contain one object, except for N159 where we target 3 YSOs. 
}\label{fig:overview} 
\end{figure*}

We restrict our target list to the brightest sources with \emph{Spitzer IRS} spectra for ground-based near-IR follow-up. 
These objects have SEDs dominated by dust emission and features in their \emph{IRS} spectra that indicate a YSO and/or a compact H~{\sc ii} region. 
We compile source classifications from the literature in Table~\ref{t:classes}. 
All of our targets fall into the \citet{kraemer2002} Group 5 -- sources with SEDs dominated by dust emission with a peak wavelength longer than 45~\micron ---  and all have strong emission features likely to trace PAH emission (5.U).  
Most sources also have strong emission lines detected in the \emph{ISO SWS} spectra (5.UE). 
\citet{seale2009} report similar classifications using a modified version of the \citet{kraemer2002} scheme. 
Following the \citet{seale2009} lexicon, sources with strong PAH emission are labeled ``P Group'' in Table~\ref{t:classes}; those that also have strong fine-structure emission lines (in addition to the PAH features) are labeled ``PE Group;'' 
sources with dominant silicate absorption are labeled ``S Group.''

More recently, 
\citet{jones2017} considered all LMC objects with staring-mode \emph{IRS} spectra within the SAGE footprint.
Using a decision-tree approach developed by \citet{woods2011}, \citet{jones2017} divide sources into 25 different classes and subclasses (see their Table~3). 
Most of our LMC targets (4/5) have atomic emission lines suggesting that an H~{\sc ii} region has already developed, though they may also have PAH features in their spectra.
Defined this way, the \textit{H~{\sc ii}} class of objects 
includes both young (compact) H~{\sc ii} regions and older (more diffuse) regions.
The only LMC source without associated H~{\sc ii} region emission in our sample is classified as a \textit{YSO2}. 
These young sources have removed some of their circumstellar material and are hot enough to destroy icy mantles such that their spectra no longer show ice absorption. 
We list the \citet{jones2017} classifications for our targets in Table~\ref{t:classes}.

Our single SMC target, the N76~YSO, shows silicate absorption in its \emph{IRS} spectrum, and thus is identified as an 'S Group' source by \citet{seale2009}.
Applying the decision-tree method, \citet{woods2011,ruffle2015} classify the source as a \textit{YSO2} (see Table~\ref{t:classes}).

\begin{table*}
\caption[Classifications]{Source classifications}
\vspace{5pt}
\centering
\vspace{3pt}
\begin{footnotesize}
\begin{tabular}{llrrrr}
\hline\hline
Target & region & K2002$^a$ & S2009$^b$ & J2017$^c$ & R2015$^d$ \\ 
\hline 
J84.703995-69.079110 & LMC/30~Dor & 5.UE & PE$^{\dagger}$ & HII & ... \\ 
J72.971176-69.391112 & LMC/N79 & ... & ... & HII & ... \\ 
J84.906182-69.769472 & LMC/N159 & 5.UE & PE & HII & ... \\ 
J84.923542-69.769963 & LMC/N159 & 5.UE & PE & YSO2 & ... \\ 
J85.018918-69.743288 & LMC/N159 & 5.UE & PE & HII & ... \\ 
J16.280250-71.995194 & SMC/N76 & ... & S$^*$ & ... & YSO-2 \\ 
\hline
\multicolumn{6}{l}{classification from $^a$\citet{kraemer2002}; $^b$\citet{seale2009}; $^c$\citet{jones2017};} \\
\multicolumn{6}{l}{                    $^d$\citet{ruffle2015} } \\
\multicolumn{6}{l}{$^{\dagger}$ strong fine-structure emission lines and PAH features in the IRS spectrum, see \citet{seale2009} } \\ 
\multicolumn{6}{l}{$^*$ spectrum dominated by 10~$\mu$m silicate absorption, see \citet{seale2009} } \\
\end{tabular} 
\end{footnotesize}
\label{t:classes}
\end{table*}


\section{Spectral Energy Distributions}\label{s:seds}
Broad source classifications for objects with multiwavelength photometry may be obtained by determining their colors \citep[e.g.,][]{bolatto2007} 
or fitting models to their spectral energy distributions \citep[SEDs; e.g.,][]{whitney2008}.
We list source luminosities obtained from SED fits in the literature in Table~\ref{t:sed_props}.
New models with more flexible parameters 
have been published since those literature values were computed \citep{robitaille2017}. 
Unlike the previous generation of models which provided uneven sampling of the parameter space and limited complexity, these new models include cold dust emission at longer wavelengths and allows the user to determine the components -- disk, envelope, cavity, and ambient medium -- included in the model fit to the SED.

\begin{figure*}
  \centering
  \includegraphics[trim=10mm 18mm 0mm 0mm,angle=0,scale=0.485]{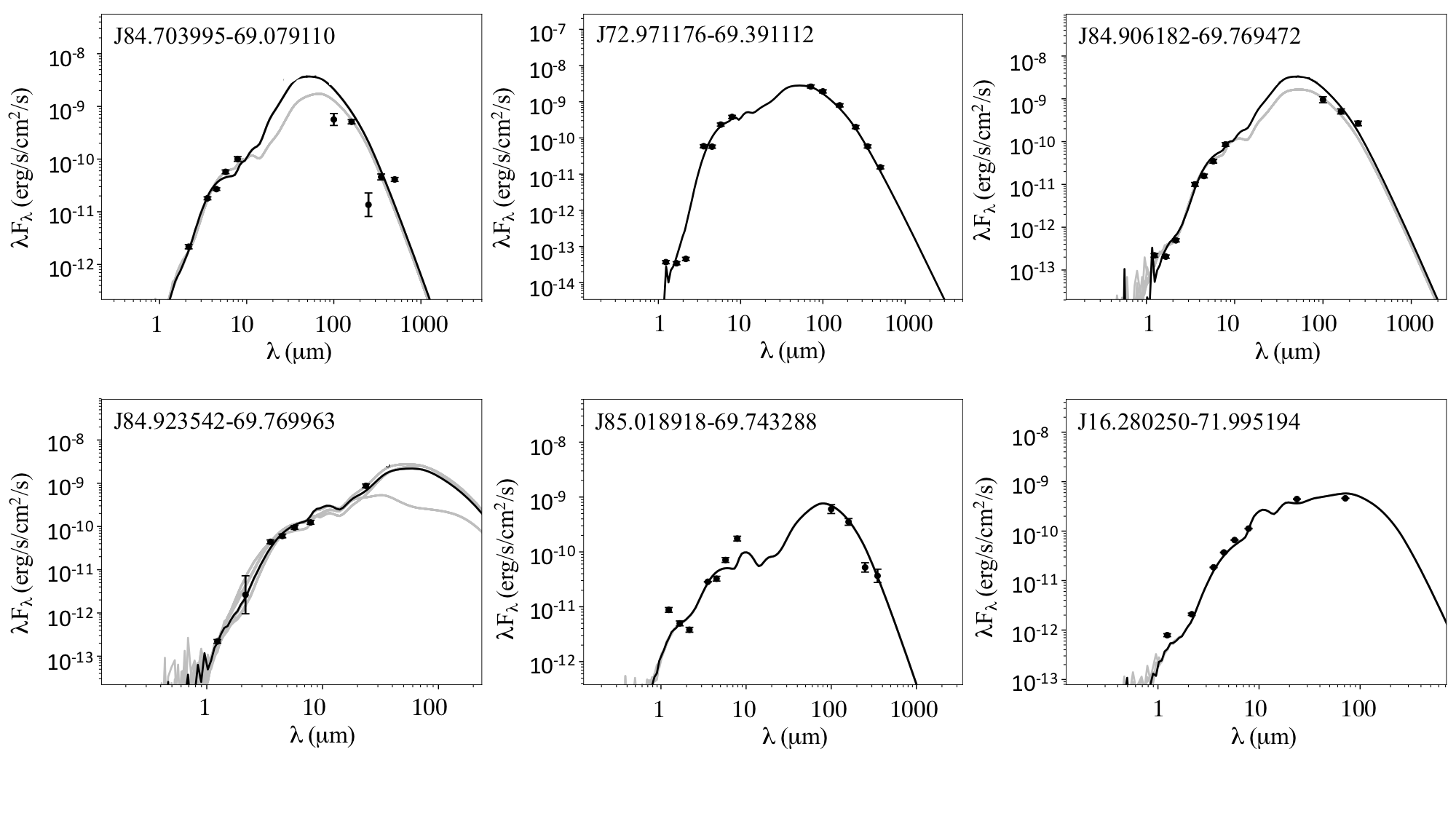} 
  \caption{SEDs of each source in our sample and the best-fitting model from \citet{robitaille2017}. 
The black line is the best-fit model and the grey lines are models with $\mathrm{\chi^{2}_{model}} - \mathrm{\chi^{2}_{best fit}} < 3$.
    Model-derived source properties are listed in Table~\ref{t:sed_props}. 
}\label{fig:SEDs} 
\end{figure*}

Applying the \citet{robitaille2017} models to the IR SEDs, we find that 
the best-fitting model for most objects in our sample
consists of a central star surrounded by 
an Ulrich envelope \citep[which has the density structure of a free-falling, rotating, and collapsing envelope, see][]{ulrich1976} with a cavity and emission from the ambient medium
\citep[the s-ubhmi model set, see Table~2 in][]{robitaille2017}.
Sources that are well-fit with these models most closely resemble the Stage I 
evolutionary classification for low-mass stars \citep[see][]{robitaille2006}. 
We list new estimates of source luminosities, temperatures, and radii in Table~\ref{t:sed_props} and show the best model fits in Figure~\ref{fig:SEDs}. 
To use these models, we have implicitly assumed that a single star dominates the flux in each source. 
Most of our sources are probably unresolved clusters, with multiple stars unresolved in the beam \citep[see, e.g.,][]{bernard2016,ward2016}.

\begin{table*}
\caption[SED]{Source properties from SED fits}
\vspace{5pt}
\centering
\vspace{3pt}
\begin{footnotesize}
\begin{tabular}{llllllll}
\hline\hline
Source & Name & log(L$_{\star}$)$^a$ & M$_{\star}$ & R$_{\star}$ & T$_{\star}$ & log(L)$^b$ & ref \\
       &      & [L$_{\odot}$] & [M$_{\odot}$] & [R$_{\odot}$] & [K] & [L$_{\odot}$] & \\ 
\hline
J84.703995-69.079110 & 30 Dor YSO & 5.70 & 24$^{\star}$ & 53.6 & 21120 & $4.83$ & \citet{nayak2016} \\
J72.971176-69.391112 & H72.97-69.39 & 6.30$^*$ & ... & ... & ... & $6.34$ & \citet{och17} \\
J84.906182-69.769472 & & 5.82 & 29$^{\dagger}$ & 60.6 & 21290 & $4.53$ & \citet{carlson2012} \\
J84.923542-69.769963 & & 5.52 & 34$^{\dagger}$ & 40.01 & 22010 & $4.92$ & \citet{nayak2018} \\
J85.018918-69.743288 & Papillon YSO & 5.07 & 50$^{\star\star}$ & 96.23 & 10960 & $5.34$ & \citet{nayak2018} \\
J16.280250-71.995194 & N76 YSO & 5.06 & ... & 22.31 & 22660 & $5.15$ & \citet{oliveira2013} \\
\hline
\multicolumn{8}{l}{$^a$ new luminosity estimates from this paper, see Section~\ref{s:seds}  } \\
\multicolumn{8}{l}{$^b$ literature values for the source luminosity } \\
\multicolumn{8}{l}{$^{\dagger}$ from \citet{chen2010} } \\
\multicolumn{8}{l}{$^{\star}$ from \citet{nayak2016} } \\
\multicolumn{8}{l}{$^{\star\star}$ from \citet{nayak2018} } \\
\multicolumn{8}{l}{$^*$ luminosity calculated by fitting two grey-bodies to the mid- to far-IR photometry (excluding the near-IR JHK points).
} \\ 
\end{tabular} 
\end{footnotesize}
\label{t:sed_props}
\end{table*}

\section{Near-IR Spectroscopy}\label{s:obs}

\begin{table*}
\caption[Observations]{FIRE observations of high-mass star-forming regions in the LMC and SMC}
\vspace{5pt}
\centering
\vspace{3pt}
\begin{footnotesize}
\begin{tabular}{llllll}
\hline\hline
Target & region & RA & DEC & K$^*$ & t$_{int}$ \\
 & & J2000 & J2000 & [mag] & [s] \\ 
\hline 
J84.703995-69.079110 & LMC/30~Dor & 05:38:48.96 & --69:04:44.8 & 14.07 & $4 \times 60$ \\
J72.971176-69.391112 & LMC/N79 & 04:51:53.08 & --69:23:28.0 & 12.39 & $4 \times 60$ \\
J84.906182-69.769472 & LMC/N159 & 05:39:37.53 & --69:46:09.8 & 12.57 & $4 \times 300$ \\
J84.923542-69.769963 & LMC/N159 & 05:39:41.89 & --69:46:11.9 & 12.16 & $4 \times 600$ \\ 
J85.018918-69.743288 & LMC/N159 & 05:40:04.54 & --69:44:35.8 & 11.73 & $4 \times 300$ \\ 
J16.280250-71.995194 & SMC/N76 & 01:05:07.26 & --71:59:42.7 & 14.12 & $5 \times 600$ \\
\hline
\multicolumn{6}{l}{$^*$ K mags from \citet{kato2007} } \\ 
\end{tabular} 
\end{footnotesize}
\label{t:obs}
\end{table*}


Near-IR spectra of 6 high-mass YSOs in the Large and Small Magellanic Clouds were obtained with the Folded-Port Infrared Echellette \citep[FIRE, ][]{sim13} on the 6.5~m Magellan/Baade telescope on 26 November 2015.
Figure~\ref{fig:overview} shows the locations of our targets within their galactic contexts. 
FIRE provides continuous wavelength coverage from $0.8-2.5$~$\mu$m with a spectral resolution of R=4800 for our chosen slit width of $0.75^{\prime\prime}$. 
Data were taken in standard ABBA sequence in windy weather with $\sim 1^{\prime\prime} - 2''$ seeing. 
Demographic information including total integration times for each target for the 6 sources targeted for near-IR spectroscopy are listed in Table~\ref{t:obs}.

Data reduction was performed using the {\sc firehose} IDL pipeline.
The software provides flat-fielding, sky subtraction, order tracing, object extraction, flux and wavelength calibration.
For most of the spectrum, wavelength calibration was performed using a ThAr lamp.
However, longward of $\sim 2.27$~$\mu$m, both ThAr and OH skylines are sparse, providing poor wavelength solutions for the reddest portions of the spectra.
In this region, 
we compute the wavelength solution directly from the data.
For J84.906182-69.769472, and J84.923542-69.769963 in N159 and H72.97-69.39, we use the detected H-Pfund lines while we fit to the H$_2$ lines for N76. 
For the YSO in 30~Dor, we used the CO (2-0) first overtone bandhead absorption. 
Few lines are detected at redder wavelengths, particularly in the Papillon YSO, making line identification difficult and the resulting correction to the wavelength solution less robust. 
We assume a systemic velocity of 262.2~km~s$^{-1}$ for the LMC and 145.6~km~s$^{-1}$ for the SMC \citep{mcconnachie2012}.

To measure line fluxes for each source, we fit each order of the spectrum with a model using a Levenberg-Markwardt least-squares fit \citep{mar09}. 
The model spectrum consists of a polynomial to represent the continuum plus a Gaussian at the position of each line.
Each order is fit separately, weighted by the 1-$\sigma$ Poisson errors computed in the {\sc firehose} reduction.
Many spectra are noisy. 
To improve the signal in these cases, we bin the data to a spectral resolution $R \approx 1200$ to improve line identification and fitting. 
All lines detected with $\geq 3 \sigma$ are listed in Table~\ref{t:detected_lines}.

\section{Spectral Features}\label{s:results}
\begin{figure*}
  \centering
    \includegraphics[trim=10mm 10mm 0mm 5mm,angle=0,scale=0.95]{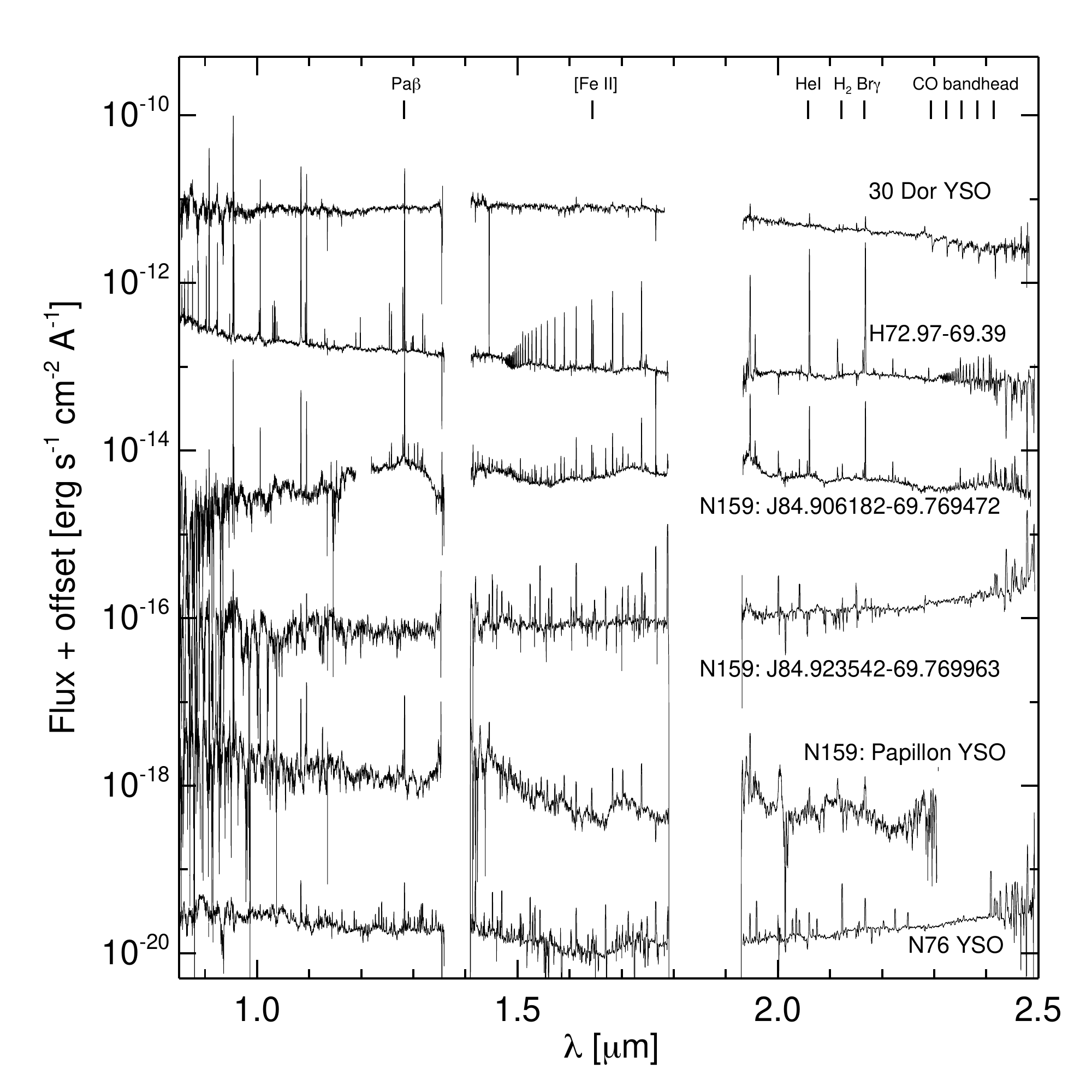}
\caption{Spectra of the high-mass star-forming regions in the LMC and SMC observed with FIRE. Source demographics are listed in Tables~\ref{t:classes},~\ref{t:sed_props}, and~\ref{t:obs} while lines detected in each object are listed in Table~\ref{t:detected_lines}. 
}\label{fig:der_data} 
\end{figure*}

Binned and smoothed spectra are shown in Figure~\ref{fig:der_data}.
The most prominent emission lines in each source are hydrogen recombination lines, although rarer and higher excitation species are seen in a few cases.
Lines detected in each source are listed in Table~\ref{t:detected_lines}. 
We briefly describe the relevant features of each object below.


\subsection{30~Dor YSO}
The most massive and luminous YSOs in the LMC reside in the R136 SSC at the heart of 30~Dor.
The YSO we target in this program lies $\sim 11$~pc outside the central SSC where strong winds and turbulence from R136 impact ongoing star formation.
J84.703995-69.079110, which we refer to as the 30~Dor YSO is the most massive YSO within the footprint of ALMA observations of 30~Dor \citep{nayak2016}.
This source remains embedded in its natal molecular gas, with a mass of $\sim 24$~M$_{\odot}$ \citep[derived by fitting the \citealt{robitaille2006} models to the SED, see][]{nayak2016}.

The spectrum of the 30~Dor YSO shows a few of the brighter hydrogen recombination lines (Pa$\beta$, Br$\gamma$) as well as He~{\sc i} emission at 2.058~$\mu$m.
We also detect high ionization lines like [S~{\sc iii}] $0.9533$~$\mu$m in the 30~Dor YSO spectrum.

The 30~Dor YSO is the only source in the sample that shows the CO bandhead in absorption. 
Molecules readily form in the cooler envelopes of evolved stars, making 
CO bandhead absorption a useful probe of their extended atmospheres 
\citep[e.g.,][]{forster-schreiber2000,bieging2002}.
In young sources, 
CO bandhead \emph{emission} has been shown to trace circumstellar disks in some high-mass young stellar objects \citep[e.g.,][]{ilee2013}.
The CO bandhead has been seen in absorption in a few lower-mass stars, particularly those with high accretion rates \citep[typically members of the FU~Orionis class of objects, see e.g.,][]{calvet1991,hk96,connelley2018}.

\subsection{Super Star Cluster H72.97-69.39 in N79}\label{s:N79}

H72.97-69.39 is a young SSC candidate located in the N79 region of the LMC. 
It is the most luminous site of star formation in the LMC, with physical characteristics that suggest it is a younger analog of the R136 SSC in the heart of 30~Dor \citep[see][]{och17}. 
ALMA observations of H72.97-69.39 suggest that the source is young as it is still associated with a significant amount of dense molecular gas. 
At the same time, the mass of ionized gas is small compared to other YSOs in the LMC, suggesting that the highest-mass sources are just beginning to ionize the surrounding gas \citep{och17}. 
Gas kinematics suggest that two colliding filaments may have stimulated the formation of the SSC.

H72.97-69.39 has the richest near-IR spectrum with several well-detected lines from multiple species.
In addition to blue continuum emission, we detect 
several hydrogen emission lines in the Paschen, Brackett, and Pfund series
as well as multiple He~{\sc i} emission lines (see Table~\ref{t:detected_lines}).  
Unlike most other sources in this sample, H72.97-69.39 also shows multiple 
[S~{\sc ii}] and [Fe~{\sc ii}] lines and a few H$_2$ emission lines. 
A more detailed analysis of this rich spectrum will be presented in a separate paper.

\subsection{YSOs in N159}

N159 is one of the richest complexes of star formation in the LMC.
Several H~{\sc ii} regions trace recent star formation \citep{chen2010} and the CO intensity is the highest of any giant molecular cloud (GMC) in the LMC with associated H~{\sc ii} regions \citep{fukui2008}. 
Three distinct clouds trace different stages of star formation in the region.
Both N159 West (N159-W) and N159 East (N159-E) contain O-type stars \citep{chen2010}, but N159 South (N159-S) does not.
Nevertheless, N159-W and N159-E have different star formation activity, suggesting that the two clouds are in different stages of their evolution. 
Protostars in N159-W appear to be younger and the majority are still found within their parental molecular gas clumps. 
Recent ALMA observations revealed the first extragalactic protostellar molecular outflows emanating from N159-W \citep{fukui2015}. 
In contrast, N159-E has more evolved protostars which have cleared out their surrounding dust and gas \citep{nayak2018}.

We observe three YSOs in N159 with FIRE: two in N159-W and one in N159-E.
In N159-W, we observed the most massive YSO that lies within the footprint of the ALMA data \citep[J84.906182-69.769472, see][]{nayak2018} as well as a YSO located at the center of two colliding filaments \citep[J84.923542-69.769963, see][]{fukui2015}. 
Both sources have masses $\sim 30$~M$_{\odot}$ \citep[derived from model fits to the SED assuming a single source dominates the luminosity, see][]{chen2010}. 
In N159-E, we target a $\sim 20$~M$_{\odot}$ protostar J85.018918-69.743288, also known as the Papillon YSO \citep{chen2010,nayak2018}. 
This source is located at the center of three colliding filaments, but its immediate environment is devoid of any molecular gas.
We describe each source and the features in its near-IR spectrum below.

\subsubsection{J84.906182-69.769472}
The spectrum of J84.906182-69.769472 is dominated by hydrogen recombination lines and He~{\sc i} emission lines.
J84.906182-69.769472 is one of three sources that shows [Fe~{\sc ii}] 1.64~$\mu$m emission. 
Near-IR [Fe~{\sc ii}] emission is often assumed to be shock excited as it is seen in protostellar jets and supernova remnants \citep[e.g.,][]{giannini2013,bruursema2014}.
However, significant FUV radiation in H~{\sc ii} regions may allow photoexcitation to dominate \citep[e.g.,][]{mouri2000_FeII}.
We consider source excitation conditions further in Section~\ref{ss:excitation}.

\subsubsection{J84.923542-69.769963} 
J84.923542-69.769963 in N159-W is one of two sources associated with a bipolar molecular outflow \citep{fukui2015}.
Indeed, the J84.923542-69.769963 YSO has the youngest evolutionary classifications of sources in our sample as a \textit{YSO2} (see Table~\ref{t:classes}). 
Near-IR continuum emission appears to rise toward redder wavelengths, as seen in younger YSOs. 
Poor telluric subtraction and the low overall S/N of the binned spectrum permits the detection of only a few hydrogen recombination lines.

\subsubsection{J85.018918-69.743288/ Papillon Nebula YSO}
The Papillon Nebula in N159E is a compact, high-excitation H~{\sc ii} region.
Like J84.923542-69.769963, the Papillon nebula lies at the intersection of molecular filaments \citep{saigo2017}.
\citet{chen2010} suggest that the central source, the Papillon YSO, is a cluster of multiple stars with a $\sim 20$~M$_{\odot}$ source dominating the flux. 
In the FIRE spectrum, we detect a few hydrogen recombination lines.

\subsection{N76 YSO} 
Our only target in the SMC is the brightest YSO in N76. 
Silicate absorption in the \emph{Spitzer IRS} spectrum indicates that the N76~YSO remains relatively embedded \citep[see Table~\ref{t:classes} and][]{oliveira2013}.
However, the absence of ice absorption indicates that the protostar has begun to heat the envelope and weak PAH emission suggests the onset of UV radiation. 
Two pure-rotational H$_2$ lines are detected in the \emph{IRS} spectrum of the N76~YSO.
\citet{ward2017} resolved this source into two components; our pointing and K~mag (see Table~\ref{t:obs}) correspond to their source 28~A which dominates the K-band emission in the region.
In addition, they find extended Br$\gamma$ and H$_2$ emission, both with velocities redshifted $\sim 5-10$~km~s$^{-1}$ relative to the continuum, that suggest the presence of an outflow. 
\citet{ward2017} also detected a second source $\sim 1^{\prime\prime}$ away, 28~B.
This second object has a featureless continuum, calling into question whether or not it is a YSO.

Our FIRE spectrum of the N76~YSO shows a few emission lines, most notably 
He~{\sc i} 1.0830~$\mu$m, 
[Fe~{\sc ii}] at 1.64~$\mu$m, and 
Br$\gamma$ at 2.16~$\mu$m.
Three H$_2$ lines -- 
H$_2$ 1-0 Q(1) at 2.4066~$\mu$m,
H$_2$ 1-0 Q(2) at 2.4134~$\mu$m, and 
H$_2$ 1-0 Q(3) at 2.4237~$\mu$m
-- appear to be marginally detected out toward $\sim 2.4$~$\mu$m, although this identification is uncertain given the poor wavelength solution.

\section{Discussion}\label{s:discussion}

We present new near-IR spectra of 5 high-mass star-forming regions in the LMC and one in the SMC.
All 6 targets have high luminosities (L$_{\star} \sim 10^5$~L$_{\odot}$), placing them well above the luminosity cutoff that \citet{cooper2013} used to identify high-mass star-forming regions in the Galaxy in the Red MSX Source (RMS) survey (L$_{\star} \sim 10^3$~L$_{\odot}$).
The FIRE spectra presented in this paper provide simultaneous coverage over $\sim 0.8 - 2.5$~$\mu$m.
Overall, hydrogen recombination lines dominate the detected emission lines, with many sources also showing H$_2$, He~{\sc i} 2.058~$\mu$m, and [Fe~{\sc ii}] emission (see Table~\ref{t:detected_lines}). 
We now examine how emission lines detected in most of our sources may be used to diagnose their physical properties.

\subsection{Comparing Source Classifications}

Emission lines detected in the near-IR spectra support the evolutionary classifications in the literature (see Section~\ref{s:seds}). 
In the near-IR, as in the \emph{IRS} spectra, we see many hydrogen recombination lines and fine structure emission lines that are commonly detected in H~{\sc ii} regions (see Table~\ref{t:detected_lines}).
This is somewhat in tension with the young evolutionary status implied by model fits to the SEDs (see Section~\ref{s:seds}). 
However, young sources emit the bulk of their emission at longer wavelengths where large beamsizes make contamination from the environment more likely, especially for distant sources like our targets in the Magellanic Clouds (at 500~$\mu$m, the Herschel beam is $36^{\prime\prime}$ corresponding to 8.7~pc at the distance of the LMC).

New ALMA data obtained with $\sim 1^{\prime\prime}$ resolution provide a better qualitative assessment of the source evolution via indicators like remnant natal material. 
Molecular gas in the immediate environs of the 30~Dor YSO, H72.97-69.39, and the two YSOs we target in N159-W suggests that these sources are younger than an object like the Papillon YSO that has largely cleared the surrounding molecular material. 
Both J84.923542-69.769963 in N159-W and the N76~YSO
show evidence for an associated outflow (see \citealt{fukui2015} and \citealt{ward2017}, respectively), consistent with these two sources having the youngest evolutionary classifications in our sample (see Table~\ref{t:classes}).  
For high-mass sources (M$>$8~M$_{\odot}$), the onset of ionizing radiation precedes the end of accretion \citep[e.g.,][]{klaassen2011,klaassen2018}. 
Therefore, the presence of emission lines from an emerging H~{\sc ii} region does not necessarily signal that accretion has finished.

Hydrogen recombination lines dominant the spectra, likely tracing emission from the associated H~{\sc ii} regions. 
Most sources also have He~{\sc i} emission (5/6).
Hard UV photons are required to excite He~{\sc i} emission, hinting that high-luminosity sources may excite the line.
However, there does not appear to be a luminosity threshold required for He~{\sc i} in Galactic high-mass YSOs \citep{cooper2013} and indeed He~{\sc i} is seen in emission around low- and intermediate-mass stars \citep[e.g.,][]{edwards2003,edwards2006,fischer2008,cauley2014,reiter2018}.  
In H~{\sc ii} regions as well as low-mass pre-main-sequence stars, photoexcitation likely dominates over collisional excitation \citep[see,][respectively]{osterbrock2006,kwan2011}. 
Given that He~{\sc i} emission is seen in young, embedded sources \citep[e.g.,][]{covey2011,connelley2014} as well as more evolved regions, it is not clearly an indicator of source mass or evolutionary stage.

We detect a larger number of near-IR emission lines from sources with the \emph{H~{\sc ii}} classification. 
The two sources in the LMC with the fewest lines detected also have the lowest luminosities (see Table~\ref{t:obs}). 
J84.923542-69.769963 has the youngest source classification of our sources in the LMC, \textit{YSO2}, and is the only source with rising continuum in the near-IR spectrum. 
Deeper integrations would provide better signal-to-noise data, allowing for more line detections, and thus more firm conclusions about the evolutionary stages of each source.

\subsection{Extinction}
To estimate the extinction to each region, we use the observed Pa$\beta$/Br$\gamma$ line ratio. 
Relative extinction between the wavelengths of the two emission lines may be computed from 
\begin{equation}
A_{rel} = -2.5 \times \log \left( \frac{ (Pa\beta / Br\gamma)_{obs} }{ (Pa\beta / Br\gamma)_{exp} } \right)
\end{equation}   
where the subscripts ``obs'' and ``exp'' denote the observed and expected flux ratios, respectively.  
For our sample, we assume Case B recombination which indicates an intrinsic flux ratio Pa$\beta$/Br$\gamma = 5.75 \pm 0.15$
appropriate for $100 < n_e < 10^4$~cm$^{-3}$ and $5000 < T_e < 10^4$~K 
\citep{storey1995}.

We adopt the extinction curves for the LMC and SMC derived by \citet{gordon2003}. 
Using values for the average extinction curve and the average $R_V$ 
derived for the LMC average and SMC bar samples 
($R_V = 3.41$ and $R_V = 2.74$, respectively), 
we compute
$\frac{ E(\lambda - V) }{ E(B - V) }$.
To convert the relative extinction to $A_V$, we use 
\begin{equation}
\frac{A_{\lambda}}{A_V} = \frac{ E(\lambda - V) }{ E(B - V) } \frac{1}{R_V} + 1
\end{equation} 
We list derived values of $E(B-V)$ and $A_V$ values in Table~\ref{t:extinction}.

Studies targeting other sight lines in the LMC and SMC report different $R_V$ values \citep[e.g., $R_V = 4.5 \pm 0.2$ for 30~Dor from][]{demarchi2016}.
Higher $R_V$ values will increase the estimated $A_V$. 
For example, extinction to the 30~Dor YSO becomes $A_V = 3.24$~mag, compared to the $A_V = 2.46$~mag reported in Table~\ref{t:extinction}. 
\citet{gordon2003} argue that their extinction curves are consistent with a continuum, reflecting the varying local impact of star formation on dust in the ISM.
Given the range of star-forming conditions sampled in this and other studies, it is possible that slightly different $R_V$ values may be appropriate for different objects.

Our targets have a range of internal extinction values.
Two of the most heavily extincted objects ($A_V \gtrsim 5$) also have the youngest evolutionary classifications (J84.923542-69.769963 and the N76~YSO, see Table~\ref{t:classes}).
Our extinction estimate for the N76~YSO is intermediate between the values determined by \citet{ward2017} who found $A_V = 0.81 \pm 0.35$~mag (optical) and $A_V = 19.5 \pm 9.3$~mag (K-band) using emission line ratios with the Galactic extinction curve \citep{ccm1989} and an $R_V = 3.1$.

We detect both the 1.26~$\mu$m and 1.64~$\mu$m lines of [Fe~{\sc ii}] in two sources, the H72.97-69.39 candidate SSC and J84.906182-69.769472. 
These two [Fe~{\sc ii}] lines originate from the same upper level, so a deviation of their observed flux compared to the intrinsic value \citep[1.49, see][]{sh06} provides an independent estimate of the reddening.
Following the same procedure as for the hydrogen lines, we find
$A_V = 3.26 \pm 1.9$ for J84.906182-69.769472,
and
$A_V = 6.24 \pm 2.6$ for H72.97-69.39.
These estimates are somewhat higher and lower, respectively, than values derived from the hydrogen lines, likely reflecting substructure in the gas 
and perhaps differences in where the emission originates within that gas.

\begin{table}
\caption[Extinction]{Extinction estimates}
\vspace{5pt}
\centering
\vspace{3pt}
\begin{footnotesize}
\begin{tabular}{llll}
\hline\hline
Name            & Pa$\beta$/Br$\gamma$ & $E(B-V)$        & $A_V^*$ \\
\hline 
30~Dor YSO           & $4.24 \pm 0.41$ & $0.72 \pm 0.16$ & $2.46 \pm 0.57$ \\ 
H72.97-69.39         & $1.56 \pm 0.01$ & $3.09 \pm 0.05$ & $10.5 \pm 0.15$ \\ 
J84.906182-69.769472 & $5.18 \pm 0.23$ & $0.25 \pm 0.05$ & $0.84 \pm 0.15$ \\ 
J84.923542-69.769963 & $1.40 \pm 0.38$ & $3.34 \pm 0.59$ & $11.4 \pm 2.05$ \\ 
Papillon YSO         & $3.20 \pm 0.31$ & $1.39 \pm 0.16$ & $4.73 \pm 0.56$ \\ 
N76 YSO              & $0.88 \pm 0.65$ & $4.43 \pm 0.11$ & $12.13 \pm 0.30$ \\ 
\hline
\multicolumn{4}{l}{$^*$ assuming $R_V$=3.41 and $R_V$=2.74 for the LMC and SMC, respectively} \\
\end{tabular} 
\end{footnotesize}
\label{t:extinction}
\end{table}

\subsection{Excitation}\label{ss:excitation}
\subsubsection{H$_2$}

Many studies of giant H~{\sc ii} regions and star-forming galaxies use wide-field, narrowband images or targeted spectroscopy to obtain diagnostic line ratios to reveal excitation conditions in the gas \citep[e.g.,][]{dale2004,riffel2013,yeh2015}. 
For example, 
the H$_2$/Br$\gamma$ ratio provides two lines that are close in wavelength and therefore relatively insensitive to extinction. 
Collisional excitation will increase the strength of the 2.12~$\mu$m H$_2$ relative to Br$\gamma$ at 2.16~$\mu$m leading to higher ratios ($ > 1$) whereas in photoexcited gas, the flux of Br$\gamma$ will far out weigh that of H$_2$, leading to a smaller ratio ($ < 0.6$).

We detect the H$_2$~1-0:S(1) line at 2.12~$\mu$m in 4/6 of the sources we observed with FIRE: 
H72.97-69.39, 
J84.906182-69.769472, J84.923542-69.769963, and 
the N76~YSO.
The H$_2$/Br$\gamma$ ratio is less than one for two of our LMC sources --
H$_2$/Br$\gamma\sim 0.01 \pm 0.001$ for H72.97-69.39, and 
H$_2$/Br$\gamma\sim 0.16 \pm 0.016$ for J84.906182-69.769472.
These small values are similar to what \citet{yeh2015} find from a large-area imaging study of 30~Dor. 
Small values in both sources suggest that fluorescence dominates over shock-excitation.

We find values slightly above one in J84.923542-69.769963 (H$_2$/Br$\gamma\sim 1.08 \pm 0.09$) and 
the N76~YSO in the SMC (H$_2$/Br$\gamma\sim 1.33 \pm 0.08$). 
These are also the youngest sources in the sample, with an evolutionary classifications of YSO-2 (unlike the H~{\sc ii} regions described above; see Table~\ref{t:classes}). 
Shocks excited by protostellar outflows from on-going star formation may contribute to the observed H$_2$ and increase the H$_2$/Br$\gamma$ ratio above the value expected for pure photoexcitation. 
\citet{ward2017} also find evidence for shock excitation in the N76~YSO from higher H$_2$/Br$\gamma$ ratios for the stellar source (28~A; H$_2$/Br$\gamma \approx 1$) and the featureless continuum souce (28~B; H$_2$/Br$\gamma > 1.3$) detected $\sim 1^{\prime\prime}$ away. 
The N76~YSO is the only source in which we detect several H$_2$ lines with $\lambda >2$~$\mu$m (see Figure~\ref{fig:der_data}).
Detailed analysis of multiple near-IR H$_2$ lines often reveal level populations intermediate between pure photoexcitation and pure shock-excitation, revealing the complex physical conditions in these regions \citep[e.g.][]{kaplan2017}. 
Similar analyses may be possible with higher quality spectra of the N76~YSO.

\subsubsection{[Fe~{\sc ii}]}

Near-IR [Fe~{\sc ii}] emission lines 
are often assumed to be collisionally-excited in shocks given their prevalence in protostellar jets and supernova remnants \citep[e.g.,][]{hollenbach1989,morel2002,giannini2013,bruursema2014}.
However, as with H$_2$, in H~{\sc ii} regions where significant FUV radiation permeates the region, [Fe~{\sc ii}] may be photoexcited. 
Line ratios like 
[Fe~{\sc ii}] 1.64~$\mu$m / Pa$\beta$ or 
[Fe~{\sc ii}] 1.64~$\mu$m / Br$\gamma$
may be used to determine the dominant excitation mechanism, as the ratio
will be much larger ($>1$) in regions where shocks play a significant role \citep[see, e.g., Table~1 in][]{labrie2006}.

We detect [Fe~{\sc ii}] emission at 1.64~$\mu$m in the same 3/6 sources in our sample where we also detect H$_2$.
Emission line ratios are $<1$ for all sources with 
[Fe~{\sc ii}] 1.64~$\mu$m / Pa$\beta$ ratio of 
$\sim 0.02 \pm 0.0005$ for H72.97-69.39, 
$\sim 0.03 \pm 0.004$ for J84.906182-69.769472, and 
$\sim 0.06 \pm 0.01$ for the N76~YSO. 
[Fe~{\sc ii}] 1.64~$\mu$m / Br$\gamma$ ratios are similar, with
$\sim 0.11 \pm 0.0003$ for H72.97-69.39,
$\sim 0.17 \pm 0.02$ for J84.906182-69.769472, and
$\sim 0.24 \pm 0.01$ for the N76~YSO. 
Photoexcitation appears to dominate in all cases,
with similar ratios to those observed in the Orion Nebula \citep[$0.009$ and $\sim 0.06$, respectively, inferred from the observations of][]{lowe1979,mouri2000_FeII}.


\begin{figure}
  \centering
  $\begin{array}{c}
    \includegraphics[trim=20mm 10mm 20mm 0mm,angle=0,scale=0.30]{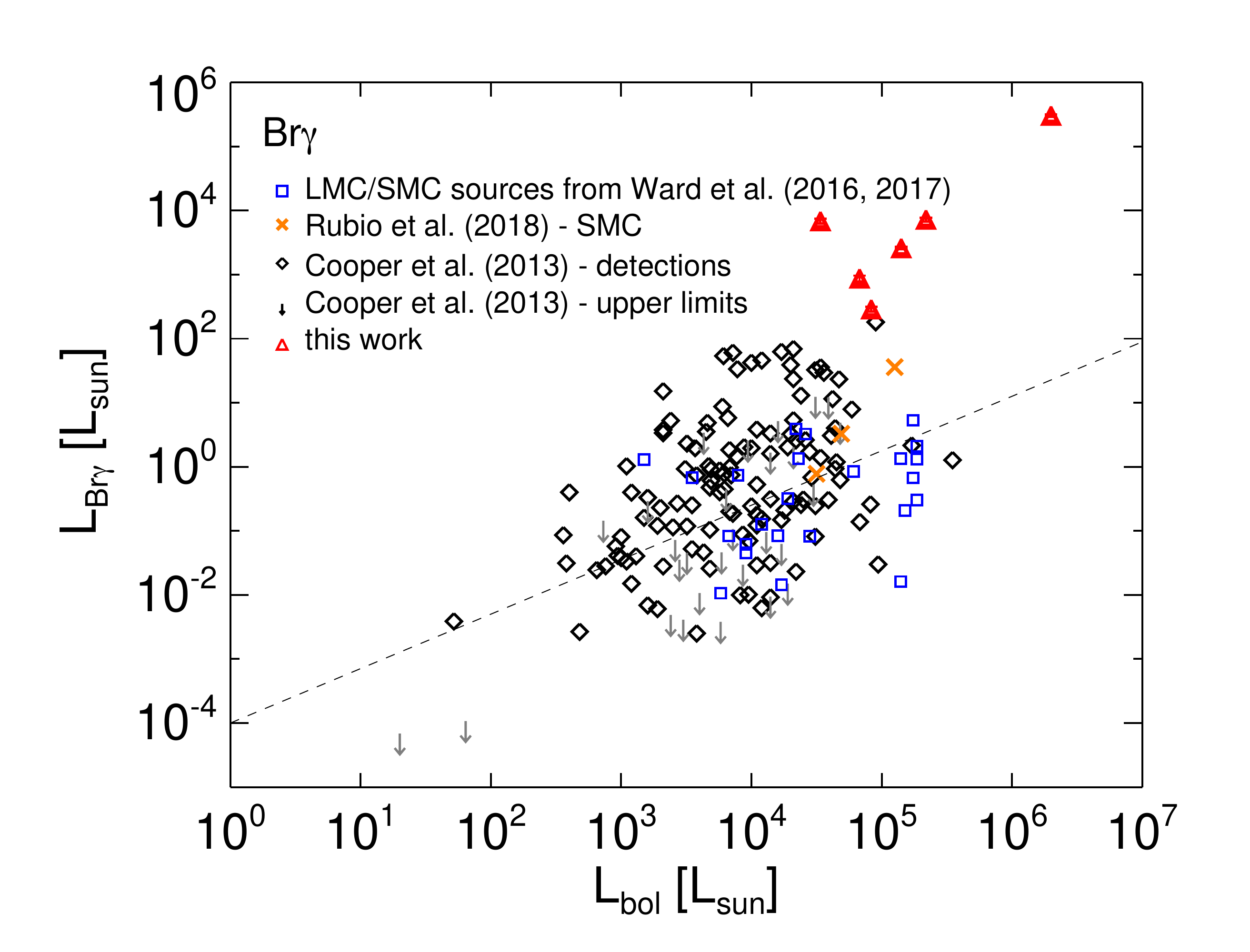} \\
    \includegraphics[trim=20mm 10mm 20mm 0mm,angle=0,scale=0.30]{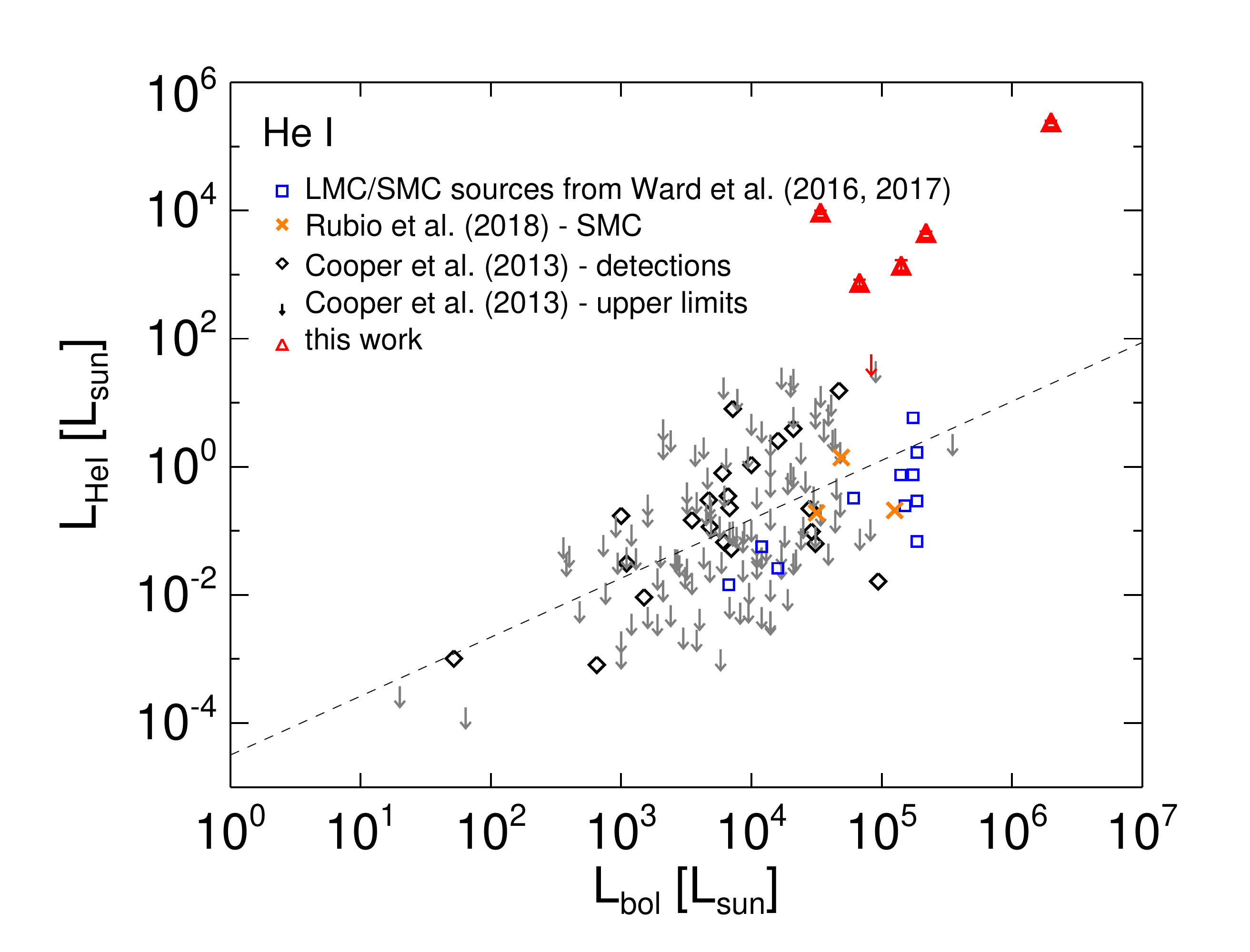} \\
    \includegraphics[trim=20mm 10mm 20mm 0mm,angle=0,scale=0.30]{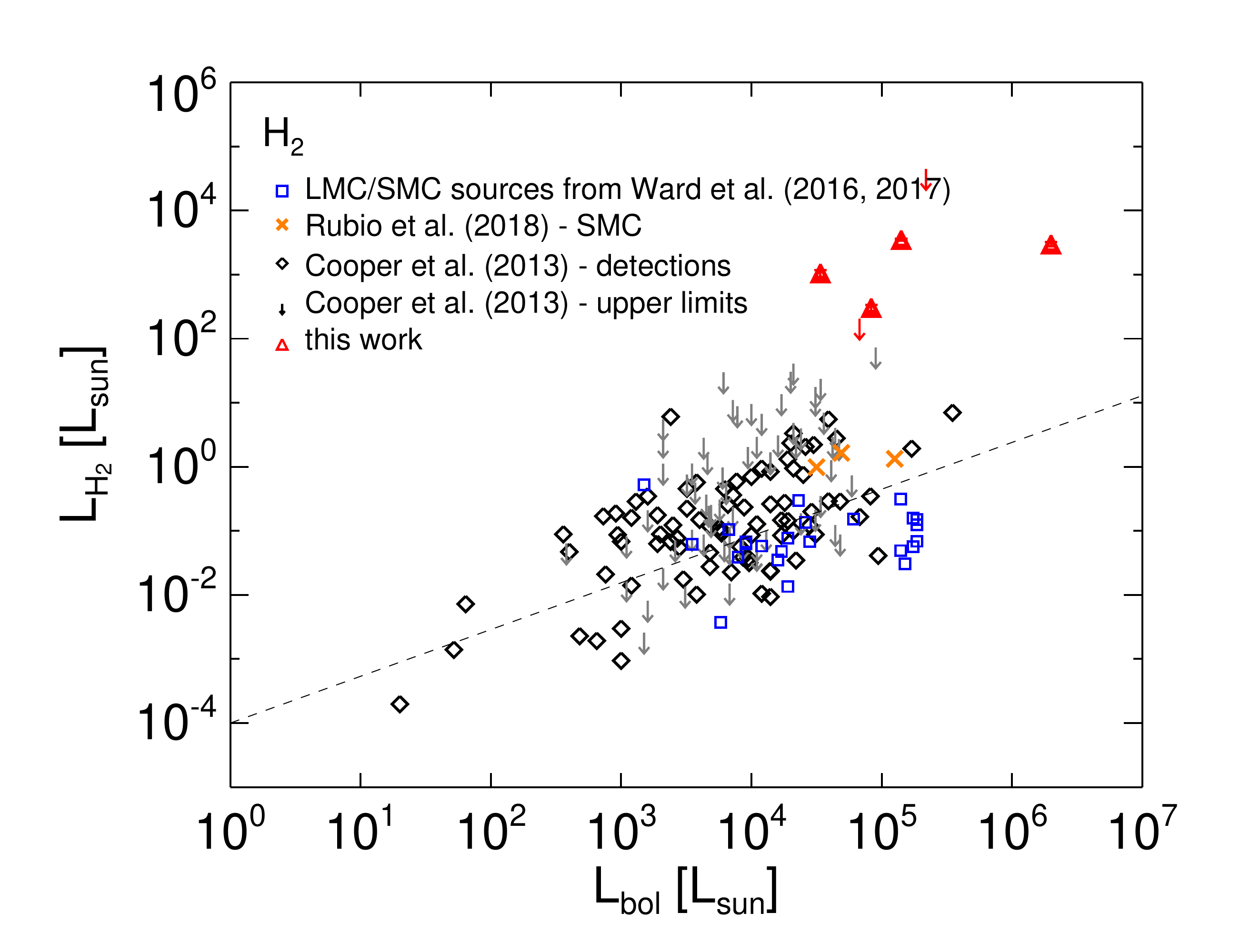} \\ 
 \end{array}$
  \caption{Line luminosities of our LMC/SMC sources (red triangles; error bars are smaller than the plot symbols and arrows indicate upper limits), LMC/SMC sources from \citet{ward2016,ward2017} (blue squares), SMC sources from \citet{rubio2018} (orange crosses), and the Galactic high-mass star-forming regions (black diamonds are detections; gray arrows are upper limits) from \citet{cooper2013} plotted as a function of bolometric luminosity. 
  }\label{fig:coop_comp} 
\end{figure}


\subsection{Comparison to other studies in Magellanic Clouds}

Much of the existing ground-based follow-up spectroscopy of Magellanic Cloud YSOs presents longer wavelength observations \citep[$\lambda \ge 3$~$\mu$m; see, e.g.,][]{maercker2005,vanloon2005,oliveira2006,shimonishi2016}.
Recently, a few studies have presented near-IR spectroscopy of high-mass YSOs, either with moderate resolution slit spectroscopy \citep[e.g.,][]{rubio2018} or higher spatial resolution integral field unit (IFU) spectroscopy \citep{ward2016,ward2017}.
We compare the line luminosities of Magellanic Cloud high-mass YSOs from \citet{ward2016,ward2017,rubio2018} as well as a sample of Galactic high-mass YSOs with our targets in Figure~\ref{fig:coop_comp} (see also Section~\ref{s:cooper_comp}). 
We specifically targeted bright sources, and indeed, our 
targets coincide with the highest luminosity LMC/SMC sources observed by other authors; line luminosities of our targets also lie above those found for sources of similar luminosity.
Better spatial resolution, particularly in IFU data \citep{ward2016,ward2017} allows for better separation between point-source and H~{\sc ii} region emission.
Indeed, nebular emission likely contaminates our observations, leading to our higher line luminosities. 
Many of our targets are also likely to be unresolved clusters (e.g., the H72.97-69.39 candidate SSC) where multiple high-mass stars contribute to the emission.

\subsection{Comparison to Galactic high-mass star-forming regions}\label{s:cooper_comp}

From more than 2000 sources discovered in the RMS survey, \citet{cooper2013} selected targets classified as probable YSOs based on multi-wavelength imaging and subject to a rough luminosity cut of $L > 10^3$~L$_{\odot}$ corresponding to
$M > 8$~M$_{\odot}$ \citep{mottram2011}.
Lines commonly detected in the spectra of those high-mass YSOs include 
Br$\gamma$, 
H$_2$,
fluorescent Fe~{\sc ii} emission at 1.68~$\mu$m, 
CO bandhead \textit{emission}, and 
He~{\sc i} at 2.058~$\mu$m.

Our targets correspond to the high-luminosity end of the \citet{cooper2013} sample and have higher line luminosities (see Figure~\ref{fig:coop_comp}).
We detect Br$\gamma$ emission in all of sources, similar to the high detection rate (75\%) reported by \citet{cooper2013}. 
We detect He~{\sc i} 2.058~$\mu$m emission in most (5/6; 83\%) of our sources, significantly more than the 15\% found among the RMS high-mass YSOs. 
Half of our sources (3/6) show both H$_2$ and [Fe~{\sc ii}]~1.64~$\mu$m emission, similar to the H$_2$ detection frequency in the RMS high-mass YSO sample (56\%; [Fe~{\sc ii}] could be separated from Brackett emission in only 11\% of the RMS sample). 
\citet{cooper2013} point to these lines as evidence of shock emission, likely signaling protostellar outflows. 
None of our [Fe~{\sc ii}]/Br$\gamma$ ratios indicate that shocks contribute significantly to the excitation. 
The median H$_2$/Br$\gamma$ ratio of the \citet{cooper2013} sample is 0.72, between the ratios computed for our sample.
Given the diversity of sources in both samples, this likely reflects a range of excitation conditions with high-mass star-forming regions.

Finally, \citet{cooper2013} report relatively high detection rates of 
fluorescent Fe~{\sc ii} emission at 1.68~$\mu$m (26\%) and
CO bandhead emission (17\%). 
Stringent excitation conditions point to an origin in circumstellar disks for both emission lines. 
We do not detect either line in emission in any of our sources.
Only one source shows the CO bandhead is seen in \emph{absorption}; we consider this peculiar object in the next Section.

\subsection{CO bandhead absorption: evidence for a disk?}\label{ss:bandhead}

We detect the CO bandhead in \textit{absorption} in the 30~Dor YSO. 
Photometric classifications sometimes confuse YSOs and evolved stars since both are red and often dust-enshrouded.
Alone, CO bandhead absorption might suggest that the 30~Dor YSO was mistakenly identified. 
However, this source has a rising continuum in the \emph{IRS} spectrum as well as PAH emission features and forbidden emission lines; these are signficantly less likely to be detected in the spectra of evolved stars.
In addition, the \emph{IRS} spectrum lacks other spectral features common to evolved stars (e.g., a 30~$\mu$m feature or C$_2$H$_2$ emission at 13.7~$\mu$mc as seen in some carbon-rich sources). 
The 30~Dor YSO is still associated with cold, molecular gas further indicating its youth \citep{nayak2016}.
One possible explanation for the coexistence of these spectral features is that the system is a binary, with one of the stars already evolving off of the main sequence. 
However, this is difficult to reconcile this hypothesis with the apparent youth of the region.

CO bandhead \emph{emission} has been observed in the spectra of young stars of both low- and high-masses. 
Several studies have successfully modeled CO bandhead emission under the assumption that it is excited in the hot ($T=2500-5000$~K), dense ($n > 10^{15}$~cm$^{-3}$) inner disk surrounding a forming star \citep[e.g.,][]{carr1989,calvet1991,glassgold2004,ilee2013,ilee2014}.
For sufficiently high accretion rates, viscous heating in an optically thick disk leads to higher temperatures in the disk midplane than in surface layers, such that the CO bandhead appears in absorption \citep[see][]{calvet1991}.
A few cases of CO bandhead absorption have been reported in the spectra of low-mass stars with high accretion rates \citep[e.g.,][]{connelley2018}. 
Such sources are often members of the FU~Orionis class of objects -- low-mass pre-main-sequence stars with sudden changes in luminosity suggesting a sudden increase in the accretion rate \citep{hk96}. 
Few emission lines tend to be detected in the spectra of outbursting sources.
This raises the intriguing possibility that the CO bandhead absorption seen in the near-IR spectrum of the 30~Dor YSO traces a disk around a high-accretion rate source in the LMC.

Detections of the CO bandhead in absorption in the spectra of higher-mass sources were reported by \citet{cooper2013} and \citet{ward2017}. 
Estimated extinctions to these three sources are somewhat higher than we estimate for the 30~Dor YSO (see Table~\ref{t:extinction}), ranging from $A_V = 4.3 \pm 24.4$~mag \citep{ward2017} to $A_V > 20$~mag \citep{cooper2013}. 
Of these three sources reported in the literature, two show H$_2$ emission; none have detected hydrogen recombination line emission. 
In contrast, multiple emission lines are seen in the near-IR spectrum of the 30~Dor YSO, although the associated H~{\sc ii} region almost certainly contributes to the emission.

The relationship between stellar effective temperature and mass accretion rate separating CO bandhead emission from absorption from \citet{calvet1991} suggests that for a source with $T > 10,000$~K (we estimate $T \gtrsim 21,000$~K for the 30~Dor YSO, see Table~\ref{t:sed_props}), the accretion rate has to be $> 10^{-5}$~M$_{\odot}$~yr$^{-1}$ for the bands to be seen in absorption.
However, it is unclear that models developed for disks around low-mass stars are appropriate for these high-mass sources. 
In particular, whether high-mass stars support optically thick, geometrically thin viscous accretion disks remains debated, with only a few detections of truly disk-like features reported in the literature \citep[e.g.,][]{kraus2010,kraus2017}. 
In addition, simulations suggest that self-gravity dominates angular momentum transport over viscosity \citep{kuiper2011}, so it is unclear whether midplane heating by accretion may be invoked in high-mass sources.

CO bandhead absorption in high-mass stars may indicate accretion even if the emission does not originate in a circumstellar disk. 
Several models for protostellar accretion suggest significant increases in the stellar radius for high accretion rates \citep[e.g.,][]{palla1992,hosokawa2009,haemmerle2016}. 
This produces a short-lived phase where the young high-mass stars is very luminous (L$_{\star} > 10^4$~L$_{\odot}$) but cool. 
In this case, the bandhead absorption may be photospheric, produced in the cool outer layers surrounding the hot nucleus of the nascent massive star. 
Such a bloated phase will be short-lived and detection of objects in this state correspondingly rare.

\section{Conclusions}\label{s:conclusions}

We present medium resolution near-IR spectra of five high-mass YSOs in the LMC and one in the SMC.
All of our objects have significant ancillary data (e.g., \emph{Spitzer IRS} spectra and molecular line observations from ALMA) that strongly suggest that they are young, high-mass stars.

Every target in our sample shows multiple hydrogen recombination lines in their spectra; 
many also show He~{\sc i} 2.058~$\mu$m. 
Emission lines detected in these near-IR spectra allow us to estimate the extinction and excitation in these sources. 
Using the ratio Pa$\beta$/Br$\gamma$, we estimate a range of internal extinctions, $1 \lesssim A_V \lesssim 12$~mag. 
Sources with the highest $A_V$ estimates tend to have the younger evolutionary classifications from other authors, indicating that they are in the early phases of formation. 
Half of the sources in our sample show [Fe~{\sc ii}] 1.64~$\mu$m emission in their spectra and 4/6 show H$_2$ 2.12~$\mu$m emission. 
Both lines are often assumed to trace shocks, but may be photoexcited in regions with strong UV fields, as in the H~{\sc ii} regions associated with most of these sources.
Ratios of [Fe~{\sc ii}] with hydrogen recombination lines indicate that photoexcitation dominates in all three sources.
However, we find 
$0.01 \lesssim$ H$_2$/Br$\gamma$ $\lesssim 1.3$, suggesting that shock contribute to the excitation of 2/4 sources.
Sources with the highest H$_2$/Br$\gamma$ ratios also have the youngest evolutionary classifications, suggesting that sources within them are still accreting and driving outflows.

We detect CO bandhead \emph{absorption} in one source, the 30~Dor YSO. 
The source is likely young as the \emph{IRS} spectra show PAH features and fine-structure emission lines typically excited in H~{\sc ii} regions and ALMA observations show that the source is still associated with its natal molecular cloud. 
However, most detections in Galactic high-mass YSOs show the CO bandhead in emission \citep[e.g.,][]{cooper2013}, 
making the origin of the CO bandhead absorption in the 30~Dor YSO somewhat mysterious.

While this study presents a limited sample of a few bright objects, it nevertheless adds to the evidence that multi-wavelength observations are essential to star formation studies in the LMC and SMC. 
With the upcoming launch of JWST, similar studies will be possible for larger samples with fainter (and lower-mass) sources. 
In the meantime, ground-based near-IR spectra like we present may be used to constrain the nature of the high-mass sources embedded and emerging from the molecular gas detected with ALMA.


\section*{Acknowledgments}
M.R. would like to thank Rob Simcoe. 
M.R. was supported by a McLaughlin Fellowship at the University of Michigan. 
M. Meixner and O. Nayak were supported by NSF grant AST-1312902.
This project has received funding from the European Union's Horizon 2020 research and innovation programme under the Marie Skl\'{l}odoska-Curie grant agreement No. 665593 awarded to the Science and Technology Facilities Council. 
This research has made use of NASA's Astrophysics Data System Bibliographic Services; the arXiv pre-print server operated by Cornell University; and the SIMBAD and VizieR databases hosted by the Strasbourg Astronomical Data Center.

\bibliographystyle{mnras}
\bibliography{bibliography_mrr}


\begin{landscape} 
\begin{table}
\caption[Observations]{near- and mid-IR photometry of FIRE targets in the LMC and SMC}
\vspace{5pt}
\centering
\vspace{3pt}
\begin{footnotesize}
\begin{tabular}{llllllllll}
\hline\hline
Region & RA & Dec & J$^*$ & H$^*$ & K$^*$ & IRAC [3.6] & IRAC [4.8] & IRAC [5.7] & IRAC [8.0]  \\
 & [J2000] & [J2000] & [mJy] & [mJy] & [mJy] & [mJy] & [mJy] & [mJy] & [mJy] \\
\hline
30~Dor & 84.7039 & -69.0791 & ... (...) & ... (...) & 1.57 (0.14) & 21.38 (1.37) & 40.42 (2.23) & 109.82 (8.92) & 264.89 (26.84) \\
H72.97-69.39 & 72.97 & -69.39 & 0.015 (0.002) & 0.019 (0.002) & 0.033 (0.003) & 70.17 (7.02) & 86.44 (8.64) & 448.37 (44.84) & 1001.90 (100.19) \\
N159 & 84.9061 & -69.7694 & 0.09 (0.01) & 0.11 (0.01) & 0.36 (0.03) & 11.93 (1.20) & 23.69 (2.40) & 66.79 (6.70) & 226.50 (22.60) \\
N159 & 84.9235 & -69.7699 & 0.09 (0.01) & ... (...) & 3.16 (3.20) & 52.55 (5.30) & 90.90 (9.00) & 182.26 (18.00) & 327.39 (33.00) \\
N159 & 85.0189 & -69.7432 & 3.63 (0.36) & 2.77 (0.28) & 2.75 (0.28) & 33.79 (0.34) & 48.87 (4.89) & 136.72 (13.67) & 460.65 (46.00) \\
N76 & 16.2803 & -71.9952 & 0.33 (0.02) & ... (...) & 1.50 (0.08) & 21.90 (0.37) & 55.40 (0.85) & 126.00 (1.31) & 295.00 (3.46) \\
\hline
\multicolumn{10}{l}{$^*$ near-IR photometry from \citet{kato2007} } \\ 
\end{tabular} 
\end{footnotesize}
\label{t:s_photometry}
\end{table}
\end{landscape} 
\begin{landscape} 
\begin{table}
\caption[Observations]{mid- and far-IR photometry of FIRE targets in the LMC and SMC}
\vspace{5pt}
\centering
\vspace{3pt}
\begin{footnotesize}
\begin{tabular}{llllllllll}
\hline\hline
Region & RA & Dec & MIPS [24] & MIPS [70] & PACS [100] & PACS [160] & SPIRE [250] & SPIRE [350] & SPIRE [500] \\
 & [J2000] & [J2000] & [mJy] & [mJy] & [mJy] & [mJy] & [mJy] & [mJy] & [mJy] \\
\hline
30~Dor & 84.7039 & -69.0791 & ... (...) & ... (...) & 19450 (5076) & 27370 (2511) & 1295 (667) & 5397 (698) & 6875 (593) \\
H72.97-69.39 & 72.97 & -69.39 & 6101.67$^*$ (610.17) & 63053.50 (6305.35) & 64607.18 (6460.72) & 42492.91 (4249.29) & 16696.69 (1669.67) & 6823.95 (682.39) & 2575.94 (257.59) \\
N159 & 84.9061 & -69.7694 & ... (...) & ... (...) & 32150 (5260) & 27330 (3704) & 22320 (2752) & ... (...) & ... (...) \\
N159 & 84.9235 & -69.7699 & 6911 (690) & ... (...) & ... (...) & ... (...) & ... (...) & ... (...) & ... (...) \\
N159 & 85.0189 & -69.7432 & ... (...) & ... (...) & 20450 (3802) & 18960 (2662) & 4429 (866.4) & 4446 (1238) & ... (...) \\
N76 & 16.2803 & -71.9952 & 3507 (22.8) & 10960 (67) & ... (...) & ... (...) & ... (...) & ... (...) & ... (...) \\
\hline
\multicolumn{10}{l}{$^*$ 22~$\mu$m emission from WISE \citep{wright2010_wise} } \\ 
\end{tabular} 
\end{footnotesize}
\label{t:l_photometry}
\end{table}
\end{landscape} 

\begin{onecolumn}
\begin{center}
\begin{footnotesize}
\begin{longtable}{llrr}
\caption[Lines]{Near-IR lines detected in LMC high-mass star-forming regions}\label{t:detected_lines} \\
\hline\hline
\vspace{5pt}
Line & $\lambda$ & Flux$^{\dagger}$ $\times 10^{-17}$ & FWHM \\ 
name & [$\mu$m] & erg~s$^{-1}$~cm$^{-2}$ & [\AA] \\ 
\endfirsthead
\hline
\multicolumn{4}{c}{\textbf{30~Dor YSO}} \\
\hline
$[$S~{\sc iii}$]$  & 0.9071   & 290.2 (13.9) & 2.1 \\
$[$S~{\sc iii}$]$  & 0.9533 & 971.5 (49.4) & 2.1 \\
Paschen 7-3  & 1.0052  &  87.8 (5.6) & 2.0 \\
He~{\sc i}   & 1.0834  & 153.5 (9.5) & 3.1 \\
Paschen 6-3  & 1.0942  & 129.6 (7.0) & 2.2 \\
Paschen 5-3  & 1.2822 &  169.9 (8.4) & 2.6 \\
He~{\sc i}   & 2.0587 &   35.6 (3.7) & 4.2 \\
Brackett 7-4 & 2.1661 &   40.1 (3.3) & 4.1 \\
CO 3-1       & 2.3227 &  -38.3 (-4.6) & 12.5 \\
CO 4-2       & 2.3525 &  -41.2 (-5.3) & 17.2 \\
CO 5-3       & 2.3829 &  -40.5 (-6.1) & 16.8 \\
\hline
\multicolumn{4}{c}{\textbf{H72.97-69.39}} \\ 
\hline
Paschen 22-3     & 0.8361 &  340.6 (31.5) & 3.1 \\ 
Paschen 21-3     & 0.8377 &  315.4 (23.8) & 2.2 \\
Paschen 20-3     & 0.8395 &  270.6 (21.2) & 1.7 \\
Paschen 19-3     & 0.8416 &  406.5 (24.8) & 2.1 \\
Paschen 18-3     & 0.8440 &  365.5 (24.3) & 1.9 \\
Paschen 16-3     & 0.8505 &  527.7 (51.1) & 2.4 \\
Paschen 15-3     & 0.8548 &  598.1 (54.0) & 2.2 \\
Paschen 13-3     & 0.8667 &  849.9 (16.5) & 1.8 \\
Paschen 9-3      & 0.9232 & 2412.6 (37.3) & 1.9 \\
$[$S~{\sc iii}$]$  & 0.9533 & 63266.5 (286.4) & 2.0 \\
Paschen 8-3      & 0.9549 & 2998.0 (140.2) & 1.9 \\
He~{\sc i}       & 1.0031 &  258.9 (11.1) & 2.7 \\
Paschen 7-3      & 1.0052 & 5161.3 (17.9) & 2.1 \\
$[$S~{\sc ii}$]$   & 1.0290 &  347.5 (32.2) & 2.2 \\
$[$S~{\sc ii}$]$   & 1.0323 &  412.4 (8.1) & 2.1 \\
$[$S~{\sc ii}$]$   & 1.0339 &  319.3 (7.7) & 2.1 \\
$[$S~{\sc ii}$]$   & 1.0373 &  146.5 (7.0) & 2.2 \\
He~{\sc i}       & 1.0833 & 23966.4 (292.5) & 2.8 \\
Paschen 6-3      & 1.0941 & 8643.5 (136.3) & 2.2 \\
He~{\sc i}       & 1.1972 &  251.0 (6.6) & 2.4 \\
He~{\sc i}       & 1.2531 &  496.2 (7.3) & 2.6 \\
$[$Fe~{\sc ii}$]$  & 1.2570 &  366.6 (77.9) & 3.8 \\ 
He~{\sc i}       & 1.2789 &  923.4 (23.0) & 2.7 \\
He~{\sc i}       & 1.2793 &  318.4 (23.5) & 3.0 \\
Paschen 5-3      & 1.2822 & 9551.9 (56.1) & 2.1 \\
O~{\sc i}        & 1.3168 &  355.9 (13.5) & 2.9 \\
Brackett 28-4    & 1.4892 &   74.8 (8.2) & 3.4 \\
Brackett 27-4    & 1.4916 &   80.6 (8.8) & 3.8 \\
Brackett 26-4    & 1.4942 &  101.1 (9.4) & 3.4 \\
Brackett 25-4    & 1.4971 &  117.0 (11.3) & 3.5 \\
Brackett 24-4    & 1.5005 &  113.8 (10.6) & 3.1 \\
Brackett 21-4    & 1.5137 &  190.1 (4.8) & 3.2 \\
Brackett 20-4    & 1.5196 &  204.1 (4.1) & 2.9 \\
Brackett 19-4    & 1.5265 &  236.4 (4.1) & 3.1 \\
$[$Fe~{\sc ii}$]$  & 1.5339 &   58.7 (3.7) & 3.0 \\
Brackett 18-4    & 1.5346 &  269.2 (4.2) & 3.2 \\
Brackett 17-4    & 1.5443 &  319.3 (4.2) & 3.3 \\
Brackett 16-4    & 1.5561 &  368.6 (5.2) & 3.3 \\
Brackett 15-4    & 1.5705 &  458.6 (15.0) & 3.6 \\
Brackett 14-4    & 1.5885 &  571.6 (9.5) & 3.7 \\
Brackett 13-4    & 1.6114 &  646.0 (10.3) & 3.1 \\
Brackett 12-4    & 1.6412 &  831.9 (6.8) & 3.4 \\
$[$Fe~{\sc ii}$]$  & 1.6440 &  411.0 (5.7) & 3.5 \\
Brackett 11-4    & 1.6811 & 1197.4 (14.8) & 3.9 \\
He~{\sc i}       & 1.7007 &  565.8 (15.8) & 3.7 \\
He~{\sc i}       & 1.7356 &   58.7 (4.8) & 3.3 \\
Brackett 10-4    & 1.7367 & 1536.6 (10.0) & 3.4 \\
Mg~{\sc i}       & 1.7456 &   86.8 (7.2) & 5.9 \\
He~{\sc i}       & 1.9548 &  378.2 (22.0) & 4.1 \\
He~{\sc i}       & 2.0587 & 4738.1 (17.0) & 4.2 \\
He~{\sc i}       & 2.1126 &  281.4 (7.5) & 5.0 \\
He~{\sc i}       & 2.1138 &   71.5 (7.5) & 5.1 \\
H$_2$ 1-0:S(1)   & 2.1218 &   59.5 (6.4) & 4.3 \\
He~{\sc i}       & 2.1613 &  153.5 (5.7) & 4.5 \\
He~{\sc i}       & 2.1622 &   57.7 (5.6) & 4.8 \\
He~{\sc i}       & 2.1647 &  296.8 (5.8) & 4.4 \\
Brackett 7-4     & 2.1661 & 6121.9 (24.6) & 4.3 \\
He~{\sc i} or [Fe~{\sc iii}$]$ & 2.2187 & 97.6 (6.0) & 4.6 \\
Pfund 29-5       & 2.3492 &  163.0 (12.5) & 8.2 \\
Pfund 28-5       & 2.3545 &   79.1 (8.6) & 4.6 \\
Pfund 27-5       & 2.3604 &   95.2 (8.5) & 4.9 \\
Pfund 26-5       & 2.3669 &  100.4 (10.1) & 6.3 \\
Pfund 25-5       & 2.3744 &  100.2 (9.3) & 4.9 \\
Pfund 24-5       & 2.3828 &  147.4 (10.1) & 4.9 \\
Pfund 23-5       & 2.3925 &  142.9 (10.3) & 5.3 \\
Pfund 22-5       & 2.4036 &  161.5 (10.0) & 5.0 \\
H$_2$ 1-0:Q(1)   & 2.4066 &  142.2 (12.1) & 5.4 \\
\hline 
\multicolumn{4}{c}{\textbf{J84.906182-69.769472 in N159}} \\ 
\hline
$[$S~{\sc iii}$]$ & 0.9533 & 1179.8 (65.2) & 2.0 \\
Paschen 8-3   & 0.9548  &   77.8 (7.9) & 2.0 \\
Paschen 7-3   & 1.0052 &  150.4 (6.3) & 2.1 \\
He~{\sc i}   & 1.0833 & 550.0 (11.0) & 3.0 \\
Paschen 6-3  & 1.0941 & 370.4 (8.15) & 2.1 \\
$[$Fe~{\sc ii}$]$ & 1.2570 & 69.6 (4.9) & 2.3 \\
Paschen 5-3   & 1.2821  & 1931.7 (28.9) & 2.5 \\
O~{\sc i}     & 1.3168 &   60.2 (4.5) & 2.8 \\
Brackett 19-4 & 1.5265 &   26.1 (1.9) & 2.9 \\
Brackett 18-4 & 1.5346 &   26.6 (2.1) & 3.3 \\
Brackett 17-4 & 1.5443 &   34.1 (2.0) & 3.1 \\
Brackett 16-4 & 1.5561 &   39.1 (2.7) & 3.3 \\
Brackett 15-4 & 1.5705 &   43.6 (3.0) & 3.5 \\
Brackett 14-4 & 1.5885 &   54.7 (4.8) & 3.6 \\
Brackett 12-4 & 1.6411 &   75.7 (8.2) & 7.7 \\
$[$Fe~{\sc ii}$]$ & 1.6439 & 61.1 (8.4) & 6.6 \\
Brackett 11-4 & 1.6811 &  188.8 (7.8) & 3.5 \\
Brackett 10-4 & 1.7367 &  301.7 (6.6) & 3.5 \\
He~{\sc i}    & 2.0587 &  531.3 (12.0) & 3.1 \\
H$_2$ 1-0:S(1) & 2.1218 &   58.2 (5.3) & 4.9 \\
He~{\sc i}    & 2.1647 &   53.0 (3.6) & 4.2 \\
Brackett 7-4  & 2.1660 &  372.6 (15.9) & 3.1 \\
$[$Fe~{\sc iii}$]$    & 2.2186 &   58.7 (6.3) & 6.3 \\
\hline 
\multicolumn{4}{c}{\textbf{J84.923542-69.769963$^*$ in N159}} \\
\hline
$[$S~{\sc iii}$]$ & 0.9071 & 16.9 (1.8) &  1.3 \\
Paschen 9-3     & 0.9230 & 44.3 (7.1) & 34.6 \\
$[$S~{\sc iii}$]$ & 0.9533 & 58.8 (5.9) &  1.6 \\
Paschen 6-3     & 1.0941 & 12.0 (1.1) &  1.2 \\
He~{\sc i}      & 1.0833 &  7.8 (1.4) &  3.4 \\
Paschen 5-3     & 1.2820 &  7.5 (1.7) &  1.4 \\
H$_2$ 1-0:S(1)  & 2.1218 &  5.6 (1.3) &  3.9 \\
Brackett 7-4    & 2.1661 &  5.3 (0.8) &  2.4 \\
\hline 
\multicolumn{4}{c}{\textbf{Papillon YSO in N159$^*$}} \\
\hline
Paschen 10-3 & 0.9016 & 87.2 (5.8) & 0.9 \\
$[$S~{\sc iii}$]$ & 0.9070 & 399.7 (31.9) & 2.1 \\
$[$S~{\sc iii}$]$ & 0.9532 & 2881.5 (3.9) & 1.0 \\
He~{\sc i}   & 1.0832 & 44.4 (4.2) & 3.4 \\
Paschen 6-3  & 1.0940 & 63.7 (2.0) & 2.5 \\
Paschen 5-3  & 1.2821 & 845.6 (72.5) & 0.9 \\
He~{\sc i}   & 2.0586 & 166.9 (2.9) & 1.6 \\
Brackett 7-4 & 2.1661 & 264.3 (11.2) & 0.6 \\
\hline 
\multicolumn{4}{c}{\textbf{N76 YSO$^*$}} \\
\hline
He~{\sc i}      & 1.0833 & 25.4 (2.4) & 6.9 \\
Paschen 5-3     & 1.2822 & 25.2 (1.6) & 5.1 \\
He~{\sc i}      & 2.0587 & 14.6 (1.0) & 11.2 \\
H$_2$ 1-0:S(1)  & 2.1218 & 37.8 (1.8) & 5.1 \\
Brackett 7-4    & 2.1662 & 28.5 (1.1) & 5.7 \\
H$_2$ 2-1:S(1)? & 2.2479 &  9.8 (1.4) & 4.5 \\
H$_2$ 1-0:Q(1)  & 2.4066 & 89.8 (1.3) & 5.1 \\
H$_2$ 1-0:Q(2)  & 2.4134 & 24.7 (1.4) & 5.8 \\
H$_2$ 1-0:Q(3)  & 2.4237 & 41.1 (1.2) & 4.9 \\
\hline
\multicolumn{4}{l}{$*$ rebinned to increase S/N} \\
\multicolumn{4}{l}{$^{\dagger}$ uncertainties shown in parentheses} \\ 
\end{longtable}
\end{footnotesize}
\end{center} 
\end{onecolumn}


\label{lastpage}
\end{document}